\documentclass{article}
\newcommand{\bfr}{\begin{flushright}}
\newcommand{\efr}{\end{flushright}}
 \usepackage{color}   
\usepackage{graphicx}
\usepackage{dcolumn}
\usepackage{bm}
\usepackage{slashed}

\begin{document}
\def\bb    #1{\hbox{\boldmath${#1}$}}
\def\agt {\raisebox{0.3ex}{$>$}\!\!\!\!\!\raisebox{-0.7ex}{$\sim$}~}
\def\alt {\raisebox{0.3ex}{$<$}\!\!\!\!\!\raisebox{-0.7ex}{$\sim$}~}

\title{Relativistic Hard-Scattering and Tsallis Fits to $p_{{}_T}$ Spectra in $pp$ Collisions at the
LHC \thanks{Presented at  the Low x Workshop, May 30 - June 4 2013, Rehovot and
Eilat, Israel}%
}
\author{Cheuk-Yin Wong\footnote{wongc@ornl.gov}
 ~~and~ Grzegorz Wilk\footnote{wilk@fuw.edu.pl}
\\
{\small $^\dagger$Physics Division, Oak Ridge National Laboratory,
 Oak Ridge, TN 37830, USA}
\\
 {\small $^\ddagger$Nuclear Centre for Nuclear Research, Warsaw 00-681, Poland 
}
\smallskip\\
}
\date{\today
}
\maketitle
\begin{abstract}
Motivated by the good Tsallis fits to the high-$p_{{}_T}$ spectra in
$pp$ collisions at the LHC, we study the relativistic hard-scattering
model and obtain an approximate analytical expression for the
differential hard-scattering cross section at $\eta\sim0$.  The
power-law behavior of the transverse spectra, in the form of $d\sigma/
dp_T^2$$ \propto$$ 1/p_T^n$, gives a power index $n$ in the range of 4.5-5.5
for jet production as predicted by pQCD, after the dependencies of the
structure functions and the running coupling constant are properly
taken into account.  The power indices for hadron production $n$ are
slightly greater than those for jet production.  
\\ ~ \\ PACS number(s): 13.87.-a, 24.85.+p, 25.75.Bh
\end{abstract}

\section{Introduction}
The spectra of the transverse momentum of produced particles in
nuclear collisions provide useful information on the dynamics of the
colliding systems.  These spectra are often described by the Tsallis
distribution \cite{Tsallis} (see also \cite{Wil12b,Wibig,View2}) in
the form
\begin{equation}
\frac{E d\sigma}{d^3p}=\frac{A}{\left ( 1+\frac{m_{{}_T}-m}{nT} \right )^n},
\end{equation}
where $A$ is a normalization constant, $T$ the `temperature'
parameter, and $n$ the power index, for produced hadrons with a mass
$m$ and transverse mass $m_T$.

For $pp$ collisions at $\sqrt{s}=$7 TeV, the $p_{{}_T}$ spectra from
0.5 GeV to 181 GeV can be described well by a power index $n=$6.60
\cite{Won12}.  The good Tsallis fits to the $p_{{}_T}$ spectra over
such a large range of $p_{{}_T}$ with only three parameters raise
interesting questions.  Why are there so few degrees of freedom in the
spectra over such a large  $p_{{}_T}$ domain? Mathematically,
the power index $n$ is related to the parameter $q$=$1+1/n$ in 
Tsallis non-extensive  statistical mechanics \cite{Tsallis}.  What is the
physical meaning of $n$?  If $n$ is related to the
power index of the parton-parton scattering law, then why is $n$$\sim$7 and not $n$$\sim$4 as predicted naively by pQCD?  In addition to the power law $1/p_{{}_T}^n$,
does the  
differential cross section  contain  other additional $p_{{}_T}$-dependent factors?
  Are the power indices for jet production
different from those for hadron production?  Do multiple parton
collisions play any role in modifying the power index $n$?
Does the hard scattering process contribute significantly to the production of low-$p_{{}_T}$ hadrons?

As the relativistic hard-scattering model is the proper model
for the high-$p_{{}_T}$ distribution of jets and hadrons in
high-energy collisions \cite{Bla74}-\cite{Won13}, we seek answers to
these questions from the relativistic hard-scattering model.


\vspace*{-0.2cm}
\section{Approximate Hard-Scattering Integral
\label{sec2}}
Approximate expressions for the hard-scattering integral were obtained
previously for simplifying cases \cite{Bla74,Won94,Won98}.  We would
like to work out an approximate analytical expression using the saddle
point method \cite{Won98,Won13}.  We consider the collision of $A$ and
$B$ in the center-of-mass frame at an energy $\sqrt{s}$ with $c$
coming out at $\eta\sim 0$ in the reaction $A+B \to c+X$.  Upon
neglecting the intrinsic transverse momentum and rest masses, the
differential cross section in the lowest-order parton-parton elastic
collisions is given by
\begin{eqnarray}
\frac{E_cd^3\sigma( AB \to c X) }{dc^3}
=\sum_{ab} \int dx_a  dx_b 
G_{a/A}(x_a) G_{b/B} (x_b) 
 \frac{E_cd^3\sigma( ab \to c X') }{dc^3},
\end{eqnarray}
where the parton-parton invariant cross section is related to
$d\sigma( ab\!\! \to\!\! c X') /dt$ by
\begin{eqnarray}
E_c \frac{d^3\sigma( ab \to c X') }{dc^3} =
 \frac{\hat s}{\pi}\frac{
  d\sigma( ab \to cX') } {dt} \delta (\hat s +\hat t +\hat u ),
\label{6}
\\
\hspace*{-2.7cm}{\rm and}~~~~~~~~~~~~~~
\hat s = (a+b)^2,~~
\hat t = (a-b)^2,~~
\hat u = (b-c)^2.~~~~~~~~~
\end{eqnarray}
We write out the momenta in the infinite momentum frame, 
\begin{eqnarray}
a&=&(x_a \frac{\sqrt{s}}{2}, ~{\bb O}_T, ~x_a \frac{\sqrt{s}}{2} ),
~~~~~
b=(x_b \frac{\sqrt{s}}{2} , ~{\bb O}_T, -x_b \frac{\sqrt{s}}{2} ),
\nonumber\\
c&=&(x_c \frac{\sqrt{s}}{2} + \frac{c_T^2}{2x_c \sqrt{s}}, ~{\bb c}_T, ~x_c \frac{\sqrt{s}}{2} - \frac{c_T^2}{2x_c \sqrt{s}}) .\nonumber
\end{eqnarray}
The light-cone variable  $x_c$ of the produced parton   $c$ is
\begin{eqnarray}
x_c=\frac{c_0+c_z}{\sqrt{s}}.
\end{eqnarray}
The constraint of $\hat s +\hat t + \hat u=0$ gives
\begin{eqnarray}
&&x_a(x_b)=x_c + \frac{c_T^2}{ (x_b 
 -\frac{  c_T^2}{x_c s})s}.
\end{eqnarray}
In this special case with $c$ coming out at
$\theta_c= 90^o$,   we have
\begin{eqnarray}
x_c&=&\frac{c_T}{\sqrt{s}},
~~~~x_a(x_b)=x_c +\frac{x_c^2}{x_b-x_c},~~~{\rm and~~~}
x_a=x_b=2x_c.
\end{eqnarray} 
We have therefore
\begin{eqnarray}
 \frac{E_c d^3\sigma( \!AB\! \!\to \! c X\!) }{dc^3}
\!\!=\!\!\sum_{ab}\!\!\! \int \!\!\! dx_b dx_a
G_{a/A}(x_a\!) G_{b/B} (x_b\!)\!
\frac{x_a x_b \delta (x_a\! -\! x_a(x_b))}{\pi (x_b-c_T^2/x_c s)} 
\frac{d\sigma( ab\! \to\! cX'\!)}{dt}.\nonumber
\end{eqnarray}
\vspace*{-0.6cm}
\begin{eqnarray}
{\rm We~
define ~}\hspace*{2.0cm}
{\cal G}_a(x_a)=x_a G_{a/A}(x_a),~~~
{\cal G}_b(x_b)=x_a G_{b/B}(x_b).~~~~~~
\end{eqnarray}
 After integrating over $x_a$, we obtain
\begin{eqnarray}
 \frac{E_Cd^3\sigma( AB \to c X) }{dc^3}
=\sum_{ab}\int  dx_b 
\frac{{\cal G}_a(x_a(x_b)) {\cal G}_b(x_b)}{\pi (x_b-c_T^2/x_c s)}
\frac{d\sigma(ab\! \to\! cX')}{dt}.
\end{eqnarray}
To integrate over $x_b$, we use the saddle point method to 
write
\begin{eqnarray}
{\cal G}_a(x_a(x_b)) {\cal G}_b(x_b)=e^{f(x_b)},
\end{eqnarray}
and expand $f(x_b)$ about its minimum at $x_{b0}$.  We obtain
\begin{eqnarray}
\int d x_b e^{f(x_b)} g(x_b)&\sim& e^{f(x_{b0})} g(x_{b0})\sqrt{
\frac{2 \pi}{-\partial ^2 f (x_{b}) /\partial x_b^2|_{x_b=x_{b0}}} }. 
\end{eqnarray}
For simplicity, we assume $G_{a/A}$ and $G_{b/B}$ to have the same
form.  At $\theta_c\sim 90^0$ in the CM system, the minimum value of
$f(x_b)$ is located at
\begin{eqnarray}
x_{b0}=x_{a0}=2x_c.
\end{eqnarray}
\vspace*{-0.5cm}
We get
\vspace*{0.3cm}
\begin{eqnarray}
E_C \frac{d^3\sigma( AB\!\! \to\!\! c X) }{dc^3}
\sim 
\sum_{ab}
B [x_{a0}G_{a/A}(x_{a0})][ x_{b0}G_{b/B}(x_{b0})]
\frac{d\sigma(ab\! \to\! cX')}{dt},
\\
{\rm where}~~~~~~~~~~~
B=\frac{1}{\pi (x_b-c_T^2/x_c s)} 
\sqrt{
\frac{2 \pi}{-\partial ^2 f (x_{b}) /\partial x_b^2|_{x_b=x_{b0}}} }. ~~~~~~~~~~~~~~~~~~~~
\end{eqnarray}
For the case of ${\cal G}_a(x_a)=x_aG_{a/A}(x_a)=A_a(1-x_a)^{g_a}$, we find
\begin{eqnarray}
E_C \frac{d^3\sigma( AB\!\! \to\!\! c X) }{dc^3}
\sim 
\sum_{ab}{ A_aA_b}
\frac{(1-x_{a0})^{g_a+\frac{1}{2}}(1-x_{b0})^{g_b+\frac{1}{2}}} 
{\sqrt{\pi g_a}\sqrt{ x_c(1-x_c)}}
\frac{d\sigma(ab\! \to\! cX')}{dt}.
\end{eqnarray}
If the basic process $ab \to cX'$ is $gg \to gg$, the cross section at
$\theta_c\sim 90^{o}$ \cite{Gas90} is
\begin{eqnarray}
\frac{d\sigma(gg\to gg) }{dt}
&\sim& 
\frac{9\pi \alpha_s^2}{16c_T^4}
\left [\frac{3}{2} \right ]^3.
\label{36}
\end{eqnarray}
If the basic process $ab \to cX'$ is $qq' \to qq'$, the cross section
at $\theta_c\sim 90^{o}$ \cite{Gas90} is
\begin{eqnarray}
\frac{d\sigma(qq' \to qq')}{dt}
&=&
\frac{4 \pi \alpha_s^2}{9c_T^4}
\frac{5}{16}.
\label{38}
\end{eqnarray}
In either case, the differential cross section varies as
$d\sigma(ab\!\!\to\!\! cX')/dt \sim \alpha_s^2/(c_T^2)^2$.
 
\section{ Parton Multiple  Scattering
\label{sec3}}
With increasing collision energies, we probe regions of smaller $x$,
where the parton density increases rapidly. The number of partons and
the total hard-scattering cross section in $pp$ collisions increases
with increasing collision energies \cite{Sjo87}.  The presence of a
large number of partons in the colliding system leads to parton
multiple scattering in which a projectile parton may make multiple
hard scattering with target partons.  It is of interest first to
explore how the power index may be affected by the multiple scattering
process.

We find that for the collision of a parton $a$ with a target of $A$
partons in sequence without centrality selection, the differential
$c_{{}_T}$ distribution is given by \cite{Won13}
\begin{eqnarray}
& &\frac{d \sigma_{H}^{(tot)} (a\to c)}{d \bb c_T}
=  A \frac{\alpha_s^2}{ c_T^4} \int d\bb b ~T(b)
\\
& &\hspace*{1.0cm}
+
\frac{A(A-1)}{2}
\frac{16\pi\alpha_s^4}{c_T^6}
\ln \{\frac{c_T}{2p_0}\}
\int d\bb b[T(b)]^2 
\nonumber\\
& &\hspace*{1.0cm}
+
\frac{A(A-1)(A-2)}{6}
\frac{936\pi^2\alpha_s^6 }{c_T^8}[ \ln \frac{c_T}{3p_0}]^2
\int d\bb b[T(b)]^3, \nonumber
\label{eq46}
\end{eqnarray}
where the terms on the right-hand side correspond to collisions of the
incident parton with one, two and three target partons, respectively.
The quantity $A$ is the integral of the parton density (structure
function) over the parton momentum fraction. This result show that
without centrality selection in minimum-biased events, the
differential cross section will be dominated by the contribution from
a single parton-parton scattering that behaves as $\alpha_s^2/c_T^4$
for the production of partons in the highest $p_T$ region, in line with
previous analyses on the multiple scattering process in
\cite{Kas87,Cal90,Gyu01}.  Multiple scatterings with $N>1$ scatterers
contribute to terms of order $\alpha_s^{2N}$
$[\ln{(C_T/Np_0)}]^{N-1}/c_T^{2+2N}$ \cite{Won13}.
\section{The Power Index in Jet Production
\label{sec4}}
From the results in the above sections, the approximate analytical
formula for hard-scattering invariant cross section $\sigma_{\rm
  inv}$, for $A+B \to c+X$ at $\eta\sim 0$, is
\begin{eqnarray}
E_c \frac{d^3\sigma( AB\!\! \to\!\! c X) }{dc^3}\propto 
 \frac{\alpha_s^2 (1-x_{a0}(c_T))^{g_a+\frac{1}{2}}(1-x_{b0}(c_T))^{g_b+\frac{1}{2}}}
{c_T^{4}\sqrt{c_T/\sqrt{s}}  \sqrt{1-x_c}}.
\label{19}
\end{eqnarray}
The power index $n$ has the value 4+1/2 in the above analytical
expression.  One can plot $\ln{ \sigma_{\rm inv}} $ as a function of
$\ln c_T$, and the slope in the linear section gives the value of $n$,
and the variation of $\ln{ \sigma_{\rm inv}} $ at high $\ln c_T$ gives
the value of $g_a$ and $g_b$.  On can also extract the value of the
power index $n(x_c)$ by considering a fixed $x_c$ and looking at two
different energies as suggested by Arleo $et~al.$ \cite{Arl10}
\vspace*{-0.2cm}
\begin{eqnarray}
\frac{\ln[\sigma_{\rm inv}(\sqrt{s_1},x_c)/\sigma_{\rm inv}(\sqrt{s_2},x_c)]}{\ln[\sqrt{s_2}/\sqrt{s_1}]} \sim n(x_c)-\frac{1}{2}.
\end{eqnarray}
We alternatively analyze the $p_{{}_T}$ spectra by using a running
coupling constant
\begin{eqnarray}
\alpha_s(Q(c_T)) = \frac{12\pi}{27 \ln(C + Q^2/\Lambda_{\rm QCD}^2)},
\label{run} 
\end{eqnarray}
where $\Lambda_{\rm QCD}$ is chosen to be 0.25 GeV to give
$\alpha_s(M_Z^2)=0.1184$ \cite{Ber12}, and the constant $C$ is chosen
to be 10, both to give $\alpha_s(Q$$\sim$$\Lambda_{\rm QCD})$ $\sim$ 0.6 in
hadron spectroscopy studies \cite{Won01} and to regularize the coupling constant for small values of $Q(c_{{}_T})$.  We identify $Q$ as
$c_{{}_T}$ and search for $n$ by writing the invariant cross section
for jet production as
\begin{eqnarray}
E_c \frac{d^3\sigma( AB\!\! \to\!\! c X) }{dc^3}=
 \frac{A\alpha_s^2(Q^2(c_T)) (1-x_{a0}(c_T))^{g_a+\frac{1}{2}}(1-x_{b0}(c_T))^{g_b+\frac{1}{2}}}
{c_T^{n} \sqrt{1-x_c}}.
\label{22}
\end{eqnarray}

\begin{figure} [h]
\hspace*{0.4cm}
\includegraphics[scale=0.43]{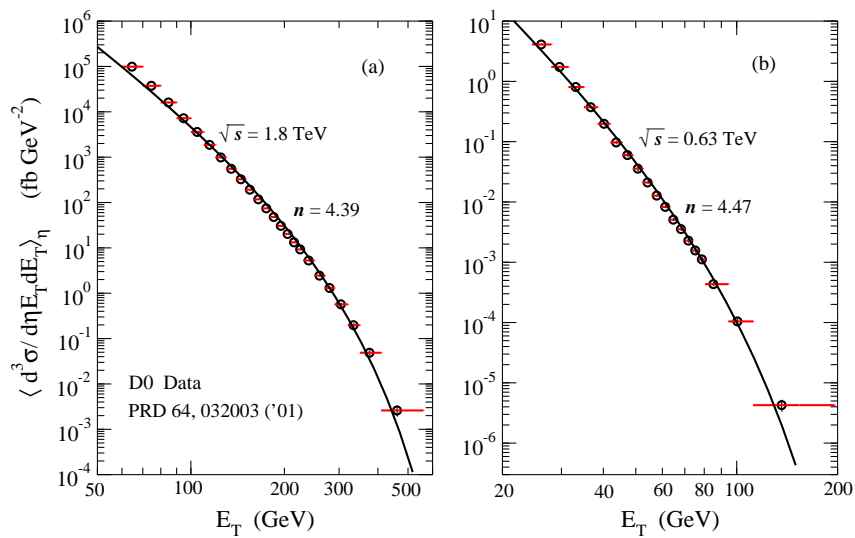}
\caption{(Color online) Comparison of the relativistic hard-scattering
  model results for jet production, Eq.\ (\ref{22}) (solid curves), with experimental
  $d\sigma/d\eta E_T dE_T$ data from the D0 Collaboration
  \cite{Abb01}, for hadron jet production within $|\eta|$$<$0.5, in
  $\bar p p$ collision at (a) $\sqrt{s}$=1.80 TeV, and (b)
  $\sqrt{s}$=0.63 TeV. }
\end{figure}

The exponential index $g_a=g_b$ for the structure function of a gluon
varies from 6 to 10 in different structure functions
\cite{Duk84,Che03,Che05}.  We shall take $g_a=6$ from \cite{Duk84}.
As shown in Fig.\ {1}, D0 $d\sigma/d\eta E_T dE_T$ data \cite{Abb01}
for hadron jet production within $|\eta|$$<$0.5 can be fitted with
$n$=4.39 for $\bar p p$ collisions at $\sqrt{s}$=1.8 TeV, and with $n$=4.47
for $\bar p p$ collisions at $\sqrt{s}$=0.630 TeV.  In other
comparisons with the ALICE data for jet production in $pp$ collisions
at $\sqrt{s}=2.76$ TeV at the LHC within $|\eta|<0.5$ \cite{Alice13}, the
power index is $n$=4.78 for $R=0.2$, and is $n$=4.98 for $R=0.4$
(Table I).  The power index is $n$=5.39, for CMS jet differential
cross section in $pp$ collisions at $\sqrt{s}=7$ TeV at the LHC within
$|\eta|<0.5$ and $R=0.5$ \cite{cms11}.  This latter $n$ value is
slightly greater than the expected value of $n=4.5$.

Except for the CMS data at 7 TeV that may need fruther re-examination,  the power indices extracted for hadron jet production as listed in Table I are in
approximate agreement with the value of $n$=4.5 in Eq.\ (\ref{19}) and
with previous analysis of Arleo $et~al.$ \cite{Arl10}, indicating the
approximate validity of the hard-scattering model for jet production
in hadron-hadron collisions, with the predominant $\alpha_s^2/c_T^4$
parton-parton differential cross section as predicted by pQCD.
\vspace*{-0.6cm}
\begin{table}[h]
\caption { The power index for jet production in
$\bar p p$ and $pp$  collisions } 
\vspace*{0.2cm}\hspace*{1.5cm}
\begin{tabular}{|c|c|c|c|c|}
\cline{1-5}
  Collaboration       &  $\sqrt{s}$ & $R$  &  $\eta$   & $n$  
  \\
\cline{1-5}
     D0             & $\bar p p$ at 1.80 TeV&  0.7  &     $|\eta|<$  0.7  & 4.39 
 \\ \hline
     D0             & $\bar p p$ at 0.63 TeV&  0.7  &  $|\eta|<$  0.7   & 4.47 
 \\ \hline
     ALICE       & $p p$ at   2.76 TeV&  0.2  & $|\eta|<$  0.5  & 4.78 
 \\ \hline
     ALICE       & $p p$ at 2.76 TeV&  0.4  &  $|\eta|<$  0.5  & 4.98 
  \\ \hline
     CMS         & $p p$ at 7 TeV&  0.5  &  $|\eta|<$ 0.5  & 5.39 
 \\ \hline
\end{tabular}
\end{table}

\section{Phenomenological Modifications for Hadron Productions}
Equation (\ref{22}) describes jet production.  To apply
Eq.\ (\ref{22}) for the case of hadron production, it is necessary to take into
account additional effects.  Jets undergoes fragmentation and
hadronization to produce the observed hadrons.  From the fragmentation
function for a parent parton jet to fragment into hadrons \cite{BKK95}, an observed hadron
$p$ of transverse momentum $p_{{}_T}$ can be estimated to arise (on
the average) from the fragmentation of a parent jet $c$ with
transverse momentum ${\bar c}_T$ \cite{Won13},
\vspace*{-0.3cm}
\begin{eqnarray}
{\bar c}_{{}_T}=2.3  p_{{}_T}.
\end{eqnarray}
Furthermore, the power law $1/p_{{}_T}^n$ appropriate for high $p_{{}_T}$
needs to be regularized for low $p_{{}_T}$.  We can regularize the factor 
$1/p_{{}_T}$ as $1/( 1+ m_{{}_T} /m_{{}_{T0}})$ by a linear dependence
on the transverse mass, $m_{{}_T}$=$\sqrt{m^2+p_T^2}$, where $m$ is
the pion mass.  With such a regularization, we examine empirically the power index $n$ in the hadron
production process $A+B \to p + X$
 by modifying Eq.\ (\ref{22}) from the case for jet
production to the case for hadron production as
\begin{eqnarray}
 \frac{d^3\sigma( AB \to p X) }{dy d{\bb p}_T}
\propto \frac{\alpha_s^2(\bar c_T)(1\!-\!x_{a0}(\bar c_T))^{g_a+1/2}(1\!-\!x_{b0}(\bar c_T))^{g_b+1/2} }
{[1+m_{T}/m_{T0}]^{n}\sqrt{1-x_c}}, 
\label{24}
\end{eqnarray}

Comparing the above equation with the hadron transverse momentum
distributions in $pp$ collisions at the LHC from the CMS \cite{CMS11},
ATLAS \cite{ATLAS}, and ALICE Collaborations \cite{ALICE} in
Fig.\ 2(b), we find $n=5.69$ and $m_{T0}=0.804$ GeV for $\sqrt{s}$=7
TeV, and $n=5.86$, and $m_{T0}=0.634$ GeV for $\sqrt{s}$=0.9 TeV
(Table II).  If we introduce $ q=1+{1}/{n}$ and $T={m_{T0}}/{q-1}$,
then we get a distribution that contains the Tsallis distribution of
Eq. (1) as a factor. The difference is the additional
$p_T$-dependencies on $\alpha_s^2(\bar c_T)$, $x_{a0}(\bar c_T)$,
$x_{b0}(\bar c_T)$, and $x_c$.
\begin{figure} [h]
\includegraphics[scale=0.52]{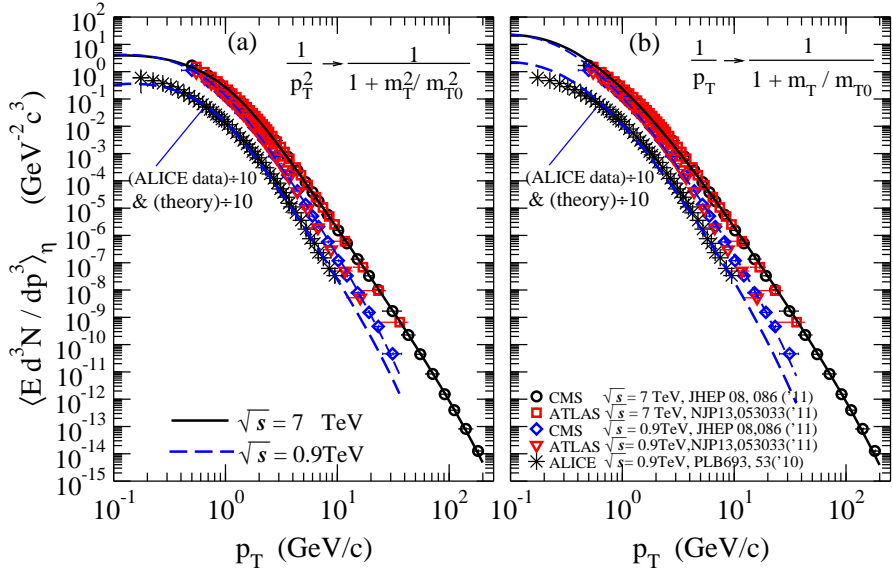}
\vspace*{-0.3cm}
\caption{(Color online) Comparison of  
 the experimental 
$\langle E_p
  d^3N/dp^3\rangle_{\eta}$ data for hadron production in $pp$
  collisions at the LHC with 
the relativistic hard-scattering model results (solid and dashed curves) (a) using
  Eq.\ (\ref{24}), with a quadratic $m_T$ dependence of the regulating
  function, and (b) using Eq.\ (\ref{25}), with a linear $m_T$ dependence of the regulating
  function.}
\end{figure}

\vspace*{-0.6cm}
Equation (\ref{24}) is not the only way we can regularize the
$1/p_{{}_T}$ behavior.  The gluon exchange propagator in the Feynman
diagrams involves the quantity $1/p_{T}^2$. We can regularize
$1/p_{T}^2$ by 
$1/(1+m_{{}_T}^2/m_{{}_{T0}}^2)$,
with
a quadratic dependence on $m_{{}_T}^2$.  We can therefore alternatively
modify Eq.\ (\ref{22}) from the case of jet production to the case for
hadron production as
\begin{eqnarray}
\frac{d^3\sigma( AB \to p X) }{dy d{\bb p}_T}
\propto \frac{ \alpha_s^2({\bar c_T})(1\!-\!x_{a0}(\bar c_T))^{g_a+1/2}(1\!-\!x_{b0}(\bar c_T))^{g_b+1/2} }
{[1+m_{T}^2/m_{T0}^2]^{n/2}\sqrt{1-x_c}}.~~~~~~~
\label{25}
\end{eqnarray}

By comparing the above equation (\ref{25}) with experimental $\langle E_p
d^3N/dp^3\rangle_{\eta}$ data for hadron production in $pp$ collisions at
the LHC from the CMS\cite{CMS11}, ATLAS\cite{ATLAS}, and ALICE
Collaborations\cite{ALICE}, we find $n$=5.45 and $m_{T0}=1.09$ GeV for
$\sqrt{s}=7$ TeV, and $n$=5.49 and $m_{T0}=0.837$ GeV for
$\sqrt{s}=0.9$ TeV (See Fig. 2(a)).  We list the parameters that
describe the $p_T$ distributions in Table II.

Comparing the results from the two different ways of expressing the
power-law behaviors, we find that the agreements of the data with the
theoretical curves are nearly the same above $p_T$$\agt $3 GeV/c, but
the theoretical results with the linear $m_T$ dependence of
Eq.\ (\ref{24}) are less than the experimental ALICE data for
$p_T$$\sim$2 GeV/c but greater than the experimental data for
$p_T$$\alt $0.5 GeV/c.  On the other hand, the quadratic $m_T^2$
expression of Eq.\ (\ref{25}), that is a more natural regularization
from the field theory point of view involving gluon propagators, leads
to a better agreement in the lower $p_T$ region.

For $pp$ collisions at the LHC, the above comparisons indicate that the
power index extracted from hadron spectra has the value of $n$$\sim$6.
The power indices for hadron production are slightly greater than the
power indices of $n$$\sim$4-5 extracted from jet transverse
differential cross sections.  Considering the differences between jets and
hadrons, we infer that the  fragmentation and showering processes
increase slightly the value of the power index $n$ of the transverse
spectra.
\vspace*{-0.7cm}
\begin{table}[h]
\caption { The power index  $n$ and $m_{T0}$ for hadron production in $pp$ collisions.
  }
\vspace*{0.2cm}\hspace*{0.5cm}
\begin{tabular}{|c|c|c|c|c|}
\cline{2-5}
   \multicolumn{1}{c}       {}        &  \multicolumn{2}{|c|} {Linear $m_T$}   & \multicolumn{2}{c|} {Quadratic $m_T^2$}  \\
   \multicolumn{1}{c}       {}        &  \multicolumn{2}{|c|} { Eq.\ (\ref{24})}   & \multicolumn{2}{c|} {Eq.\ (\ref{25}) }  \\
\cline{2-5}
       \multicolumn{1}{c|} {}               & $\sqrt{s}$=7TeV&  $\sqrt{s}$=0.9TeV  & $\sqrt{s}$=7TeV  & $\sqrt{s}$=0.9TeV  
 \\ \hline
$n$    & 5.69 &  5.86 &   5.45&  5.49   \\ \hline
$m_{T0}$ (GeV)  & 0.804 & 0.634 &   1.09  &  0.837   \\ \hline
\end{tabular}
\end{table}
\vspace*{-0.5cm}
\section{Conclusions and Discussions
\label{sec6}}

Using the saddle point integration method, we obtained an approximate
analytical expression for the differential hard-scattering cross
section at $\eta\sim 0$ with a power index of 4+1/2 in pQCD, in
approximate agreement with the experimental data for jet production.
The power indices for hadron production is empirically found to be
slightly greater than those for jet production.

With the regularization of both the power law $1/p_T^n$ and the running coupling constant $\alpha_s$ for small values of $p_{{}_T}$, the hard-scattering model has been applied to extrapolate to hadron production in the low-$p_{{}_T}$ region in Fig.\ 2.  It should be noted that  in this low-$p_T$ region, the hard-scattering cross section will be 
modified by the intrinsic $p_T$ of the partons
\cite{Won98}, the parton recombination effects \cite{Hwa03}, and the
small $x$ dependencies of the structure functions.  Nevertheless, the
extrapolation of the hard-scattering results to the low-$p_T$ region
as obtained here in Fig.\ 2 indicates indeed that the hard-scattering process can
contribute substantially to the production of particles in the
low-$p_T$ region.

Regarding the Tsallis distribution which motivates the present
investigation, we can conclude that the successes of representing the
transverse spectra at high-$p_T$ by a Tsallis distribution arises from
(i) the simple power-law behavior of the parton-parton scattering
cross section, $\alpha_s^2/c_T^4$, with a power index of $4$, (ii) the
few number of the degrees of freedom in the hard-scattering model, and
(iii) the power index of 4 that is not significantly modified by the
multiple scattering process at high $p_T$ in minimum biased
measurements.  The $\alpha_s^2/p_T^4$ power law lays the foundation
for Tsallis/Hegedorn-type transverse momentum distributions, and the
few degrees of freedom in the Tsallis distribution is a reflection the
few degrees of freedom in the underlying hard-scattering model.  There
are additional $p_T$ dependencies due to the parton structure function,
the running coupling constant, and the parton momentum integration,
which lead to a slightly larger power index.  Furthermore, in going
from the parton measurements in terms of jets to hadron measurements
in terms of fragmented hadron products, there are additional showering
and fragmentation processes which give rise to a greater value of the
power index.  The Tsallis distribution is flexible enough to adjust
the power index to accommodate the different and changing environment,
yielding a non-statistical description of the distribution.

Because of its non-statistical nature, the parameters in a Tsallis
distribution can only be supplied and suggested from non-statistical
means, such as the QCD basic parton-parton scattering power index and
the QCD multiple scattering shadowing effects.  It also is limited in
its application to the transverse degree of freedom, as there is no
way to generalize the Tsallis parameters across the three-dimensional
space from transverse to longitudinal coordinates.  For a more
fundamental description, it is necessary to turn to the basic parton
model for answers.  The underlying relativistic hard-scattering model
has a greater range of applications and a stronger theoretical
foundation.

\vspace*{0.2cm} \centerline{\bf Acknowledgment} 
\vspace*{0.2cm}
The research was supported in part by the Division of Nuclear Physics,
U.S. Department of Energy (CYW) and by the Ministry of Science and
Higher Education under contract DPN/N97/CERN/2009 (GW).




\vspace*{-0.3cm}
\bibliographystyle{apsrev4-1}


\end{document}